# Stego-Image Generator (SIG) - Building Steganography Image Database

Thiyagarajan P[#1], Aghila G[#2], Prasanna Venkatesan V[#3]
[#]*CDBR-SSE Lab Department of Computer Science, Pondicherry University, Puducherry 605 014*
[1]thiyagu.phd@gmail.com, [2]aghilaa@gmail.com
[3]prasanna_v@yahoo.com

**Abstract.** Any Universal Steganalysis algorithm developed should be tested with various stego-images to prove its efficiency. This work is aimed to build the stego-image database which is obtained by implementing various RGB Least Significant Bit Steganographic algorithms. Though there are many stego-images sources available on the internet it lacks in the information such as how many rows has been infected by the steganography algorithms, how many bits have been modified and which channel has been affected. These parameters are important for Steganalysis algorithms and it helps to rate its efficiency. Images are chosen from board categories such as animals, nature, person to produce variety of Stego-Image.

*Keywords*— Image Steganography, Steganalysis,

## 1 Introduction

Steganography is the practice of concealing the very presence of message during communication [1]. Like two sides of the coin, Steganography has both advantages and disadvantages. It depends on the person how he uses it for example it is in the hands of the scientist he may use it for the military purpose or if it is in the hands of the terrorists he will make use of steganography to snatch the attack plan among his team members and communicate via internet. In the later case it is more important to detect the image which is used by terrorists. The technique which is used to identify the images which is hands of terrorist that contains the secret message is called as Steganalysis. Our work aims to built variety of stego-image database which will be helpful for steganalyst to test their Steganalysis algorithms. Section 2 gives outline about Steganography, Section 3 gives brief outline about Steganalysis, Section 4 outlines about the Image Steganography tools and the need for RGB StegoTool, Section 5 describes the Architecture of RGB Stego Tool and experimental results, Section 6 concludes the paper.

## 2 Steganography

Steganography can be best explained through prisoner's problem Alice wishes to send a secret message to Bob by hiding information in a clean image. The stego image (clean image + secret message) passes through Wendy (a warden) who inspects it to determine if there is anything suspicious about it. Wendy could perform one or several scan to decide if the message from Alice to Bob contains any secret information. If the decision is negative then Wendy forwards the message to Bob—Wendy acts as a passive warden. On the other hand, Wendy can take a conservative approach and modify all the messages from Alice to Bob irrespective of whether any information is hidden by Alice or not.

In this case, Wendy is called an active warden. Of course, Wendy will have constraints such as the maximum allowable distortion when modifying the message etc. For example, if the clean messages are digital images, then Wendy cannot modify the stego message to an extent that perceptually significant distortions are induced. Fig 1 represents the pictorial representation of how message is embedded and extracted in steganography.

LSB methods are most commonly used steganography techniques to embed in cover image. Least significant bit of some or all of the bytes inside an image is changed to a bit of the secret message. When using a 24-bit RGB Color image, a bit of each of the red, green and blue color components can be used. In other words, one pixel can store 3 bits of secret message.

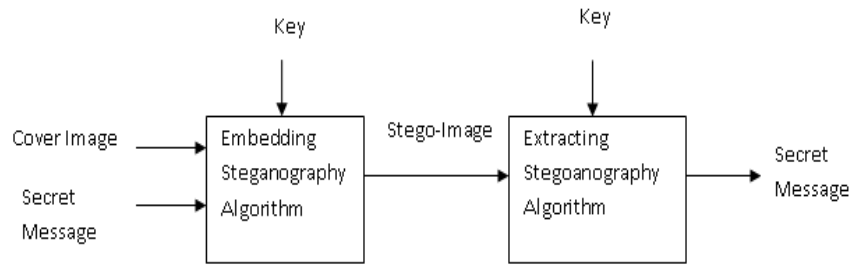

Fig. 1 Steganography Process Diagram

Components

- Secret Message  - The message to be embedded
- Cover Image – An image in which Secret Message will be embedded.
- Stego Image - Cover image that contain embedded message.
- Key – Additional data that is needed for embedding and extracting process.
- Embedding Steganography Algorithm -   Steganography Algorithm used to embed secret message with cover image.
- Extracting Steganography Algorithm - Inverse function of embedding, in which it is used to extract the embedded message (secret message) from stego image.

### 3.  Steganalysis

Steganography can be applied in numerous field such as authentication, secret communication in military, banking etc., however it also depends on the person who is using. There are strong indications that steganography has been used for planning criminal activities [2]. In this way, it is important to detect the existence of hidden messages in digital files. As with cryptography and cryptanalysis, Steganalysis is defined as the art and science of breaking the security of steganography systems. The goal of steganography is to conceal the existence of a secret message. While the goal of Steganalysis is to detect that a certain file contains embedded data. The stegosystem can be extended to include scenarios for different attacks [4] similar to the attacks on cryptographic systems. In general, extraction of the secret message could be a harder problem than mere detection. The challenges of steganalysis are listed below

a) To get Stego-Image Database
b) To test the Steganalysis algorithm against different payload Stego-Images to check its robustness

c) To test the Steganalysis algorithm from various categories of images such as animals, fruits, natural scene etc.,
d) Identification of embedding algorithm
e) Detection of presence of hidden message in cover signal
f) Estimation of embedded message length
g) Prediction of location of hidden message bits
h) Estimation of secret key used in the embedding algorithm
i) Estimation of parameter of embedding algorithm
j) Extraction of hidden message

In this paper we addresses the first three issues of Steganalysis by generating different stego-images using SIG (Stego-Image Generator).

## 4. Stego-Image Tools Reported in Literature Survey

We have surveyed 15 Stego-images generation tools. Detailed survey of the below tools have been made and they have been classified according to the format of stego-image it produces, availability of code. Table 1 gives details about the various stego-image generation tools where the code is available. Table 2 gives the detail about the Stego-image generation tool where the code is not available but its executable is available.

Table 1: Image Steganography Tools with Source Code

| S.No | Name of Image Steganography Tool | Format | Availability of Code |
|---|---|---|---|
| 1 | Blind slide | BMP | Yes |
| 2 | Camera Shy | JPEG | Yes |
| 3 | Hide4PGP | BMP | Yes |
| 4 | JP Hide and Seek | JPEG | Yes |
| 5 | Jsteg Jpeg | JPEG | Yes |
| 6 | Mandelsteg | GIF | Yes |
| 7 | Steghide | BMP | Yes |
| 8 | wbStego | BMP | Yes |

Table 2: Image Steganography Tools without Source Code

| S.No | Name of Image Steganography Tool | Format | Availability of Code |
|---|---|---|---|
| 1 | Camouflage | PNG | No |
| 2 | Hide & Seek | BMP,GIF | No |
| 3 | S-Tools | BMP | No |
| 4 | Steganos | BMP | No |
| 5 | StegMark | BMP,GIF,PNG | No |
| 6 | Invisible Secrets | BMP,JPEG,GIF | No |
| 7 | Info Stego | BMP,JPEG,PNG | No |

## 5. Architecture of SIG

Though there are many tools available for stego-image generation but it has following disadvantages
a) Lack of getting user desired inputs such as channels in which the changes to be incorporated, number of bits to be affected, number of rows to be affected.
b) Proper Classification of Stego-images with respect to number of rows affected
c) Proper Classification of Stego-images with respect to number of bits changed
d) Proper Classification of Stego-images with respect to number of channels changed

To overcome the above disadvantage a new architecture was proposed for Stego-Image Generator (SIG) tool.

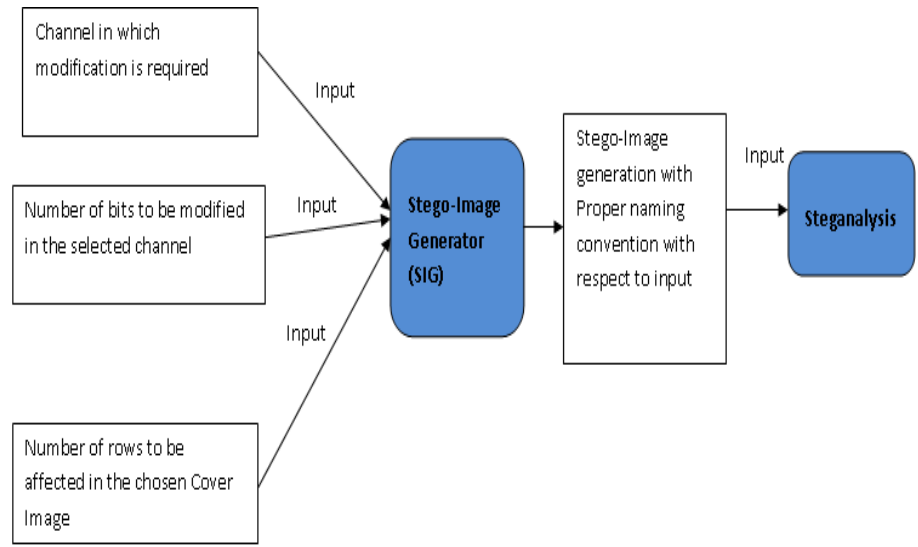

Fig. 2 Architecture of Stego-Image Generator

This architecture was designed specifically for the RGB based color Images and it was implemented in MATLAB 7. This tool can generate 63 different kinds of Stego-Images from a single Image. The calculation is shown below

In 7 different ways RGB channels can be chosen such as R, G, B, RG, RB, GB and RGB. In 3 different ways the Rows can be selected such as 5, 10, 20 rows. In 3 different ways the Least Significant Bits are changed such as 1,3 or 4 bits. Reason for choosing maximum 20 rows is that in steganography message communicated will not be of much length which can be accommodated with in 20 rows. In Stego-Image maximum of 3 to 4 bits are changed in the LSB of every pixel. The complete calculation of 63 Stego-Image obtained from a single Cover-Image is shown below

Number of ways in which RGB Channels can be chosen = 7
Number of ways in which the rows can be chosen = 3
Number of ways in which the number of bits that need to changed to be chosen = 3

Total number of Stego-Images generated for single Cover Image = 7*3*3 = 63

> **Algorithm Used in Stego-Image Generator**
>
> ***Input:*** *Cover Image, Input_Channel, No of Rows to be affected (Input_Rows), No of bits to be change (Input_Bits)*
>
> ***Output:*** *Stego-Image*
>
> ***Algorithm:***
>
> *For 1 to Input_Row*
>
>     *For 1 to last_col*
>
>         *Get the Input_channel and change the*
>
>         *LSB_Bits according to the Input_Bits*
>
>     *End*
>
> *End*
>
> *Save the generated Stego_Imae with the meaningful naming convention such as*
>
> *"filename_Inputbits_InputRows_InputChannel"*

The above LSB algorithm used in Stego-Image Generator covers most of the RGB Steganography algorithm such as Pixel Indicator High capacity Technique Algorithm, RGB Intensity Image stego Algorithm etc. These 63 RGB Stego-Images generated by SIG forms the superset therefore any RGB Steganalysis developed can extensively test with SIG to test their robustness of the algorithm. So far 50 Cover-Images from different categories have been given as input to SIG. Images from different categories have been chosen to have various color combination in the Cover-Image. Some of the sample input is shown in table 3. SIG database contains 3150 Stego-Images.

Table 3: Input Image Category for SIG Tool

| S.No | Image Name | Category |
|---|---|---|
| 1 | Lotus.bmp | Flora |
| 2 | Monkey | Fauna |
| 3 | Baby | People |
| 4 | Sea | Natural |
| 5 | Cupcakes | Eatables |
| 6 | Tajmahal | Building |

Screen shots

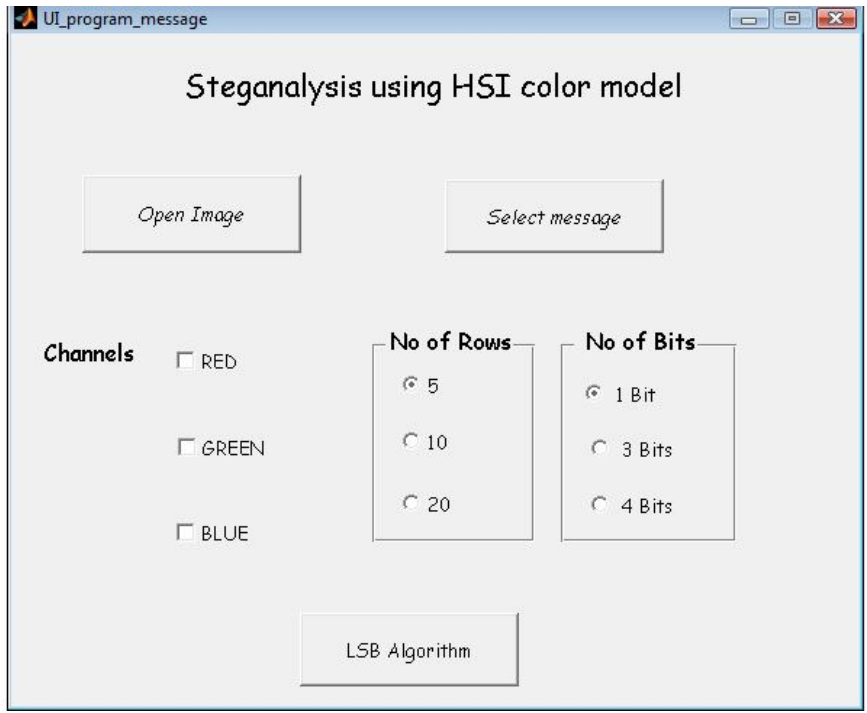

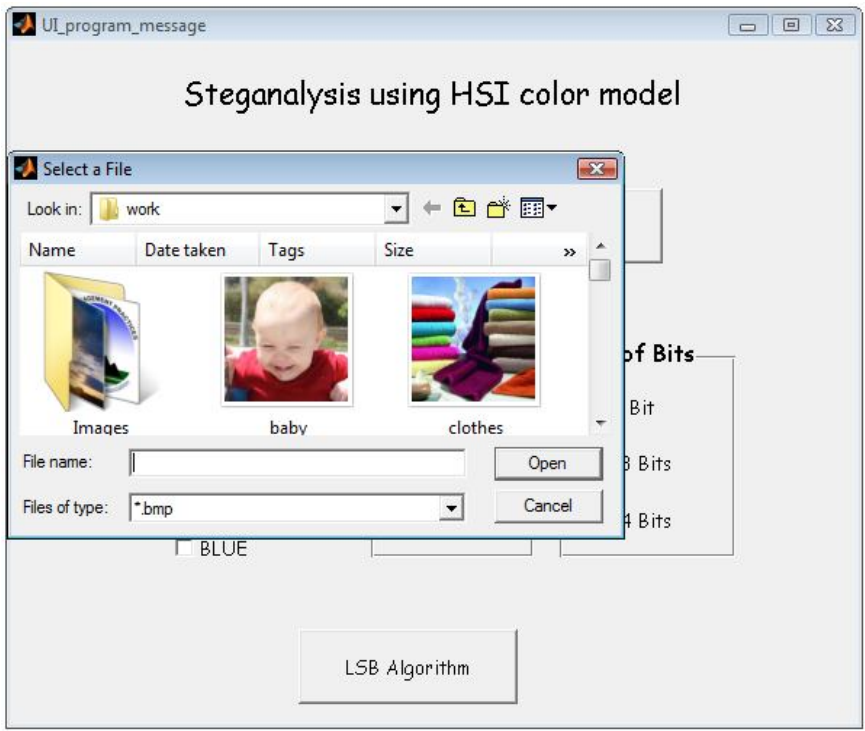

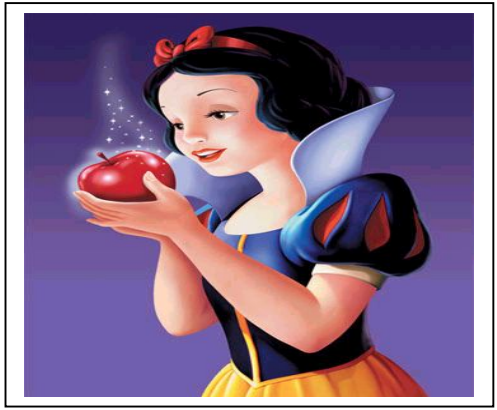 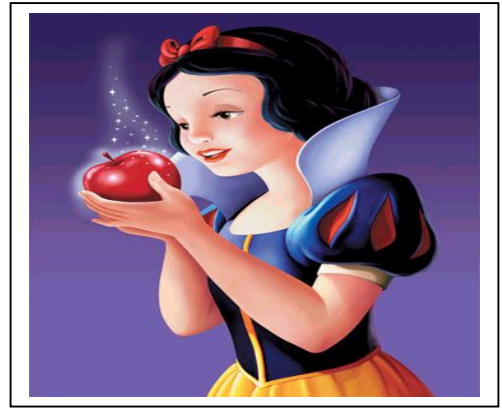

Fig 3: Cover Image						Fig 4: Stego-Image

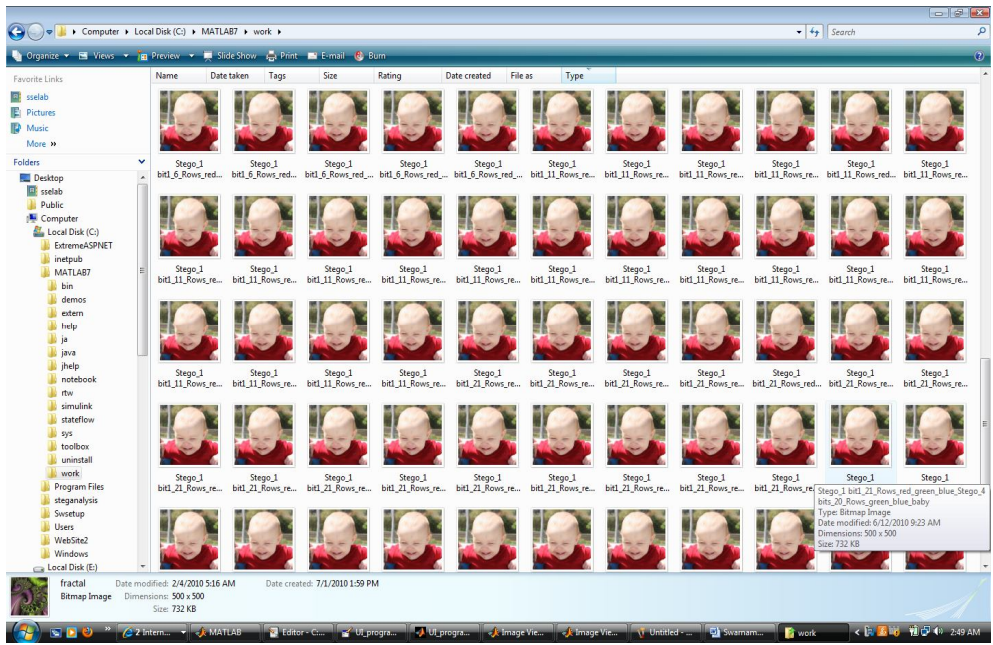

Fig 5: SIG Image Database (Stego-Images with proper naming convention)

## 6. Comparison of Stego-Image Generator with Other Tools

There are numerous tools available either in free or licensed version in the Internet for the production of Stego-Images. Mother of all Steganography algorithms in spatial domain is Least Significant Bit (LSB) algorithm. Our proposed SIG tool uses the LSB algorithm combined with choice in selecting the no of LSB bits to be replaced, choice in selecting the colour channel, choice in selecting the no of rows to be affected. In table 4 the proposed SIG tool is compared with various Stego-Tools against various parameters.

| S.No | Tools | Input such as Channel, Number of LSB to be replaced in Cover Image | Format | Number of Stego-Images Produced on per Cover Image |
|---|---|---|---|---|
| 1 | S-Tools | No | BMP | 1 |
| 2 | Camera Shy | No | JPEG | 1 |
| 3 | Steganos | No | BMP | 1 |
| 4 | Steghide | No | BMP | 1 |
| 5 | Info Stego | No | BMP,JPEG,PNG | 1 |
| 6 | SIG (Stego-Image Generator) | Yes | All format expect compressed image format | 63 |

Stego-Image Generator (SIG) produces stego-images of all types expect compressed images type. All the Stego-Tool reported in the literature survey produces 1 stego-Image for the given cover Images. SIG Tool produces 63 different types of Stego-Images.

## 7. Conclusion

In this paper 15 different steganography tools have been surveyed and the area in which it lacks has been identified. To overcome the limitations in the existing system the SIG architecture has been proposed and implemented. This tool is very useful for forensic investigators and security agencies across the globe that is constantly analysing the Internet to detect communication that involves steganography. In the future we would test all these tools to develop robust real time steganalysis algorithm that can filter Steganography images which is used by terrorists for illegal activities.

## References

[1] Berghel, H. and O'Gorman, L. Protecting ownership rights through digital watermarks. IEEE Computer. 29, 7 (July 1996), 101–103.

[2] Andreas Westfeld, Andreas Pfitzmann, Attacks on Steganographic Systems, Proceedings of the Third International Workshop on Information Hiding, p.61-76, September 29-October 01, 1999

[3] D Artz, "Digital Steganography: Hiding Data within Data". IEEE Internet Computing: Spotlight pages 75-80 May-June 2001

[4] Johnson, N. and Jajodia, S. Exploring steganography: Seeing the unseen. IEEE Computer. 31, 2 (Feb. 1998), 26–34.